\documentclass[12pt]{iopart}

\usepackage{graphicx}

\begin{document}

\title{Possible features of galactic halo with electric field and observational constraints}

\author{Koushik Chakraborty}
\address{Department of Physics,
Government Training College, Hooghly 712103, West Bengal,
India\\koushik@iucaa.ernet.in}

\author{Farook Rahaman}
\address{Department of Mathematics,
Jadavpur University, Kolkata 700032, West Bengal,
India\\rahaman@iucaa.ernet.in}

\author{Saibal Ray}
\address{Department of Physics,
Government College of Engineering and Ceramic Technology, Kolkata
700010, West Bengal, India\\saibal@iucaa.ernet.in}

\author{Arka Nandi}
\address{Department of Physics, Government
Training College, Hooghly 712103, West Bengal,
India\\arkanandi001@gmail.com}

\author{Nasarul Islam}
\address{Department of Mathematics,
Danga High Madrasa, Kolkata 700130, West Bengal,
India\\nasaiitk@gmail.com}

\begin{abstract}
Observed rotational curves of neutral hydrogen clouds strongly
support the fact that galactic halo contains huge amount of
nonluminous matter, the so called gravitational dark matter. The
nature of dark matter is a point of debate among the researchers.
Recent observations reported the presence of ions of O, S, C, Si
etc in the galactic halo and intergalactic medium. This supports
the possibility of existence of electric field in the galactic
halo region. We therefore propose a model of galactic halo
considering this electric field arising due to charged particles
as one of the inputs for the background spacetime metric.
Considering dark matter as an anisotropic fluid we obtain the
expressions for energy density and pressure of dark matter there
and consequently the equation of state of dark matter. Various
other aspects of the solutions are also analyzed along with a
critical comparison with and constraints of different
observational evidences.
\end{abstract}


\maketitle

\section{Introduction}
The flat rotation curves of the neutral hydrogen clouds in the
halos of the galaxies reveal that there must be huge amount of
nonluminous matter in the galactic halo region
\cite{Oort1932,Zwicky1933,Zwicky1937,Freeman1970,Roberts1973,Ostriker1974,Einasto1974,Rubin1978,Sofue2001}.
Researchers often assume these nonluminous matters whose existence
is understood by their gravitational effects only, to be
gravitational dark matter. Therefore the assumption of dark matter
provides the gravitational field needed to match the observed
galactic flat rotation curves in galaxy clusters.

However, there are a number of proposals for the dark matter
component. New exotic particles predicted by supersymmetry
\cite{Jugman1996}, massive neutrinos, ordinary Jupiter-like
objects etc are some of the proposed candidates for dark matter.
Of the many proposed candidates for dark matter, the standard cold
dark matter (SCDM) paradigm is possibly the most favored one
\cite{Efstathiou1990,Pope2004}. However, $\Lambda$-CDM model is
also a favored candidate for dark matter
\cite{Tegmark2004a,Tegmark2004b}. Rahaman et al.
\cite{Rahaman2008} and Nandi et al. \cite{Nandi2009} provided a
summary of the alternative theories (e.g. scalar-tensor and brane
world models) regarding dark matter (also see the Refs.
\cite{Rahaman2012,Rahaman2014} for details of dark matter in the
galactic halo). In this paper, we have modelled dark matter as
anisotropic fluid. Our conjecture is inspired by the work of
Rahaman et al. \cite{Rahaman2010} who successfully considered dark
matter as perfect fluid.

One can think of existence of charge and hence electrostatic field
in the galactic halo in a following threefold ways: (1) a galaxy
is an assemblage of stars ($\sim 10^{11}~M_{\odot}$), (2) stars
are connected with interstellar medium (ISM), and (3) galaxies are
connected with intergalactic medium (IGM). As such, as in the
first and second possibilities, the effect of charge and the
possibility of holding charge by the stars are not unavailable in
the literature
\cite{Pannekoek1922,Rosseland1924,Eddington1926,Cowling1929}. In
connection to such study Shvartsman \cite{Shvartsman1971} argued
that while astrophysical systems are usually thought to be
electrically neutral, this may not be always true in the real
situation. Basically, his analysis invokes the exchange processes
between stars and the surrounding ISM. Bally and Harrison
\cite{bally} has shown in their paper that the gravitationally
bound systems like stars, galaxies etc., carry positive charge and
being freely expandable, the intergalactic medium holds the
electrons. They pointed out that giant galaxies have potential
difference $\sim 10^{3}$ V between centre and surface. Recently
Neslu{\u{s}}an \cite{Neslusan2001} has reported about the
existence of a global electrostatic field of the sun and other
normal stars. He has formulated a general charge-mass relation,
$q_r = [2\pi \epsilon_0 G(m_p - m_e)m_r]/q$, where $\epsilon_0$ is
the permittivity of the vacuum, $G$ is the gravitational constant,
$m_r$ is the stellar mass in the sphere of radius $r$, $m_p$ and
$m_e$ are respectively the mass of the proton and the electron
having charge $q$, while $q_r$ is the global electrostatic charge
inside the star. The charge can be found out to be $q_r =
77.043~m_r$, with $q_r$ in Coulomb and $m_r$ in solar masses
corresponding to an ideally quiet, perfectly spherical,
non-rotating star. de Diego et al. \cite{deDiego} have explained
the physical mechanism that is responsible for an accreting
blackhole to be positively charged. The electrons being dominated
by the radiation pressure of the accretion disc may reach the disc
corona. A large fraction of these electrons may actually be
Compton scattered from the infalling plasma. Thus a charged
blackhole at the centre of the galaxy might be responsible for
giving rise to an electric field in the galactic halo region and
in the galaxy as a whole.

Yet another possibility of existence of charge and hence
electrostatic field in the galactic halo is as follows:\\ Very
recently Howk and Consiglio \cite{Howk2012} have presented direct
measure of the ionization fractions of several elements in the
galactic warm ionized medium. They have pointed out the existence
of ions like S~II, S~III, S~IV, C~IV, N~IV, O~VI in the galactic
halo region. Other recent observations also indicate that IGM,
which is the medium of ionized gas filling the space between
galaxies, contain considerable amount of ionized metals such as
O~VII, O~VIII etc. \cite{Meiksin2009}. In fact, several authors
have reported absorption lines of C~III, C~IV, N~V, Si~III, Si~IV
in the observed spectra of background Quasi Stellar Objects (QSOs)
\cite{Tripp2000,Savage2003,Yao2005,Danforth2006,Lehner2006,Yao2007,Tripp2008,Yao2009,Yao2010}.
In an interesting paper Fraternalli et al. \cite{Fraternali2006}
have pointed out that low angular momentum material from the IGM
should accrete to the gaseous halos. Since galactic halos, as well
as, the intergalactic spaces contain charged metal ions, one may
predict the existence of an electric field in gaseous halos. In
this connection one can note that long ago Hall \cite{Hall1986}
proposed models with millicharged matter in the galactic halo. His
argument was based on strongly interacting particles which are the
galactic halo dark matter and carry electric charge of order
$10^{-3}~q$. Quite recently the same millicharged matter have been
considered by Berezhiani, Dolgov and Tkachev \cite{Berezhiani2013}
for creation of electric current in the galactic halo (also see
the Ref. \cite{Harari2010}). However, in the present paper we
propose a model of galactic halo considering the existence of an
electric field in addition to the dark matter, which is taken as
an anisotropic fluid. The study of anisotropic fluid distribution
under General Theory of Relativity is an active field of research
\cite{lake,herrera1,herrera2,nguyen1}. Sgr\'{o}, Paz and
Merch\'{a}n \cite{sgro} have given a formalism for deriving
cross-correlation function, which is dependent on the directions
of the axis of halo shape. They have extended the classical model
of halo for triaxial nature of halo profile. A model of self
gravitating anisotropic system in post Newtonian approximation is
also available in the literature~\cite{nguyen2}. The density
profile which Nguyen and Pedraza \cite{nguyen2} have obtained was
found to be suitable for modelling galaxies and dark matter halos.

A common trend in the theoretical studies of galactic halo is to
assume a distribution function for mass in the galactic halo
region and investigate the shape of the galactic halo. However,
most of the observational evidences are provided on the shape of
the halos and rotational curves of neutral hydrogen cloud. So we
tried to find out the density and other parameters on the basis of
observed shape of the halo and predicted gravitational potential
in the halo region.

Although galactic halo is mostly believed to be spherical, recent
numerical simulations based on observational data predict
considerable departure from our common notion. Navaro, Frenk and
White \cite{Navaro1996,Navaro1997} in their simulations found
unexpected scaling pattern in galactic halos. Later on many higher
resolution simulations confirmed their prediction
\cite{Fukushinge1997,Fukushinge2001,Jing2000}. Jing and Suto
\cite{Jing2002} provided non-spherical simulations of twelve
galactic halos. In their numerical simulation based on the
observation of globular cluster tidal stream $NGC~5466$, Lux, Lake
and Johnson \cite{Lux2012} have found that halo of milky way
galaxy is either oblate or fully triaxial in shape. In this paper
we have considered a paraboloidal spacetime in the galactic halo
region. The metric considered here was originally used by Finch
and Skea \cite{Finch1989} for modelling of relativistic stars.
Later on Jotania and Tikekar \cite{Jotania2006}, in their paper,
explored the physical nature of the metric.

\section{Geometry of the galactic halo region}
The recent simulations report non-spherical shape of the galactic
dark matter halo
\cite{Navaro1996,Navaro1997,Fukushinge1997,Fukushinge2001,Jing2000,Jing2002,Lux2012}.
Though the exact shape of dark halo is yet to be understood fully,
we take paraboloidal Finch and Skea \cite{Finch1989} metric to
describe the spacetime in galactic halo region. Thus the spacetime
of the outer region of the galactic halo is assumed to be
described by the metric  of the following form
\begin{equation}\label{line1}
ds^2=-e^{\nu(r)}dt^2 + \left(1 + \frac{r^2}{R^2}\right) dr^2+r^2
d\Omega^2,
\end{equation}
where $d\Omega^2=d\theta^2+{sin}^2\theta\, d\phi^2$ and $R$ is the
parameter responsible for the geometry of the galactic halo. We
are using here geometrized units in which $G=c=1$.

The general energy momentum tensor is
\begin{equation}
T_\nu^\mu=  ( \rho + p_r)u^{\mu}u_{\nu} - p_r g^{\mu}_{\nu}+
            (p_t -p_r )\eta^{\mu}\eta_{\nu} \label{eq:emten}
\end{equation}
with $$ u^{\mu}u_{\mu} = - \eta^{\mu}\eta_{\mu} = -1. $$

The Einstein-Maxwell equations for the line element (\ref{line1})
are
\begin{equation}
\frac{1}{R^2}\left(3 + \frac{r^2}{R^2}\right)\left(1 + \frac{r^2}{R^2}\right)^{-2} = 8 \pi \rho + F, \label{eq:lam}
\end{equation}

\begin{equation}
\left(1 + \frac{r^2}{R^2}\right)^{-1}\left[\frac{\nu^\prime}{r} + \frac{1}{r^2}\right] - \frac{1}{r^2}=
8\pi p_r - F, \label{eq:nu}
\end{equation}

\begin{eqnarray}
\left(1 + \frac{r^2}{R^2}\right)^{-1}
\left[\frac{\nu^{\prime\prime}}{2} + \frac{\nu^{\prime}}{2r} +
\frac{{\nu^{\prime}}^2}{4}\right] - \frac{1}{R^2}\left(1 +
\frac{r^2}{R^2}\right)^{-2}\left[1 + \frac{\nu^\prime r}{2}\right]
=8\pi p_t
 + F,
\label{eq:tan}
\end{eqnarray}
where $F$ is the electric field ($F > 0$) with proper charge
density $\sigma $ and the equation of state along the radial
direction is
\begin{equation}
p_r = \omega \rho, \label{eq:eos}
\end{equation}
where $\omega$ is the equation of state parameter and the
Maxwell's equation can be written as
\begin{equation}
(r^2F)^{\prime} = 4\pi r^2 \sigma e^{\frac{\lambda}{2}}.
\label{eq:elec1}
\end{equation}
Equation (\ref{eq:elec1}) may  be expressed in the form
\begin{equation}
F(r) = \frac{1}{r^2}\int_0^r 4\pi r^2 \sigma
e^{\frac{\lambda}{2}}dr.  \label{eq:elec2}
\end{equation}

\subsection{Calculation of metric potentials from galactic rotation curves}
The neutral hydrogen clouds in the galactic halo region may be
considered as test particles. To derive the tangential velocity of
the neutral hydrogen clouds let us take the gravitational field
inside the halo to be characterized by the line element
(\ref{line1}). Therefore the Lagrangian for a test particle
travelling on the spacetime (\ref{line1}) can be given by
\begin{equation}\label{E:Lagrangian2}
2\mathcal{L}=-e^{\nu(r)}\dot{t}^{2}+e^{\lambda(r)}\dot{r}^{2}+r^{2}\dot{\Omega}^{2},
\end{equation}
where as usual $d \dot{\Omega}^2=d
\dot{\theta}^2+{sin}^2\theta\,d\dot{\phi}^2$. The over dot denotes
differentiation with respect to affine parameter $s$.

From (\ref{E:Lagrangian2}), the geodesic equation  for material
particle can be written as
\begin{equation}\label{E:motion}
\dot{r}^{2}+V(r)=0,
\end{equation}
which now yields the potential
\begin{equation}\label{E:V2}
V(r)=-\left(1 + \frac{r^2}{R^2}\right)^{-1} \left(e^{-\nu(r)}E^{2}-\frac{L^{2}}{r^{2}}-1\right).
\end{equation}

Here the conserved quantities $E$ and $L$, the energy and total
momentum, respectively are given by $E=-e^{\nu(r)}\dot{t}$,
$L_{\theta}=r^2\dot{\theta}$, and
$L_{\phi}=r^{2}{sin}^{2}\theta\,\dot{\phi}$. So the square of the
total angular momentum is $L^{2}=
{L_{\theta}}^{2}+(L_{\phi}/\sin\theta)^{2}$.

Following Landau and Lifshitz \cite{Landau1998} and Nucamendi,
Salgado and Sudarsky \cite{Nucamendi2001}, one can find the
tangential velocity of the test particle in the form:
\begin{equation}\label{E:tangential1}
(v^{\phi})^2=\frac{e^{\nu}L^2}{r^2E^2}=\frac{1}{2}\,r\,\nu'.
\end{equation}

This expression can be integrated to yield the metric coefficient
in the region up to which the tangential velocity is constant and
can be given by
\begin{equation}\label{E:tangential2}
e^{\nu}=B_0 r^l,
\end{equation}
where $B_0$ is an integration constant and $l=2(v^{\phi})^2$.

Equivalently, the line element (\ref{line1}) in the constant
tangential velocity regions can be written as
\begin{equation}
ds^2=-B_0 r^l dt^2+\left(1 +
\frac{r^2}{R^2}\right)dr^2+r^2d\Omega^2.\label{E:line2}
\end{equation}

\subsection{Physical solutions of the Field equations}
From the Eqs. (\ref{eq:lam}) and (\ref{eq:eos}) we get the
expression for density in the following form:
\begin{equation}\label{E:rho}
\rho = \frac{1}{8 \pi (1+\omega)} \left[ \frac{1}{R^{2}} \left(3 +
\frac{r^2}{R^2}\right) \left(1 + \frac{r^2}{R^2}\right)^{-2} +
\left(1 + \frac{r^2}{R^2}\right)^{-1}\left(\frac{l}{r^2} +
\frac{1}{r^2}\right) - \frac{1}{r^2}\right].
\end{equation}

\begin{figure}
    \centering
        \includegraphics[scale=.3]{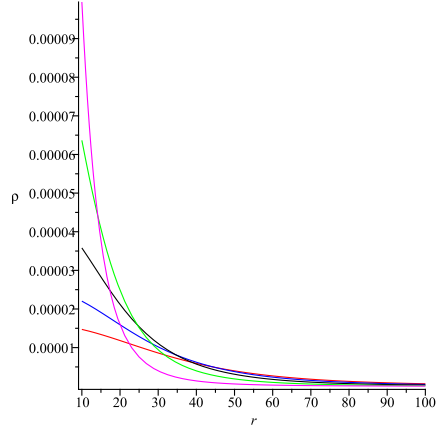}
    \caption{Plot for the variation of $\rho$ vs $r$ (in kpc).
    Here $R = 50, 40, 30, 20, 10$ for Red, Blue, Black, Green and Magenta respectively. }
    \label{1}
\end{figure}

The plots (Fig.~1) for variation of density with distance show
interesting characteristics for the distribution of mass in
galactic halo region. It is influenced by the value of $R$. The
parameter $R$ indicates a scaling factor. The plots show that the
density in the galactic halo region varies as inverse square of
the distance. The observed flat rotation curves in the galactic
halo region predict similar behavior in the distribution of masses
in the region \cite{Rubin1970,Roberts1973,Ostriker1974,Rubin1978}.

In a similar way, from the Eqs. (\ref{eq:nu}), (\ref{eq:tan}) and
(\ref{eq:eos}) we get the expressions respectively for radial and
tangential pressures in the following forms:
\begin{equation}
p_r = \frac{\omega}{8 \pi (1+\omega)} \left[ \frac{1}{R^{2}}
\left(3 + \frac{r^2}{R^2}\right) \left(1 +
\frac{r^2}{R^2}\right)^{-2} + \left(1 +
\frac{r^2}{R^2}\right)^{-1}\left(\frac{l}{r^2} +
\frac{1}{r^2}\right) - \frac{1}{r^2}\right],
\end{equation}

and

\begin{eqnarray}
p_{t} = \frac{1}{8 \pi}\left[\left(1 + \frac{r^2}{R^2}\right)^{-1} \left\{\frac{l^2}{4 r^2} - \frac{1}{1 +
 \omega} \left(\frac{l +1}{r^2}\right)\right\} \right. \nonumber
\\ - \left. \frac{1}{R^2}\left(1 +
 \frac{r^2}{R^2}\right)^{-2}
\left\{1 + \frac{l}{2} + \frac{\omega}{1 + \omega} \left(3 +
\frac{r^2}{R^2}\right)\right\} +\frac{1}{r^2 ( 1 + \omega
)}\right].
\end{eqnarray}

\begin{figure}
    \centering
        \includegraphics[scale=.3]{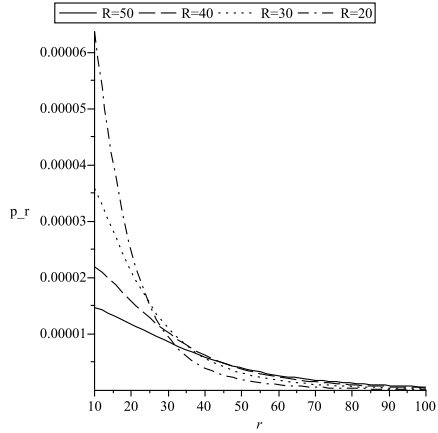}
    \caption{Plot for the variation of $p_r$  vs $r$ (in kpc). }
    \label{2}
\end{figure}

Since pressure directly depends on the density of the galactic
fluid, its value must be very small in galactic halo region. From
Fig. 2 it may be noted that the radial pressure decreases sharply
with distance from an initial large value and becomes very small
after $50$ kpc. Thus our model predicts that the value of radial
pressure must be very small in the galactic halo region. Though
cold dark matter (CDM) models predict dark matter to be
constituted of weakly interacting massive particles (WIMPS) which
offers zero pressure, Bharadwaj and Kar \cite{Bharadwaj2003}
successfully modelled galactic halo considering dark matter having
different pressures in radial and transverse directions. Recently
Harko and Lobo \cite{Harko2012} also proposed a model of galactic
halo considering characteristic pressure of dark matter.

\begin{figure}
    \centering
        \includegraphics[scale=.3]{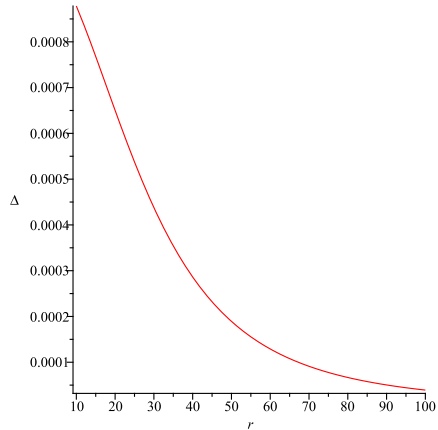}
    \caption{\small Plot for the variation of $\Delta = (p_{t} -p_{r})$  vs $r~(R = 40)$ in galactic halo region. }
    \label{3}
\end{figure}

\begin{figure}
    \centering
        \includegraphics[scale=.3]{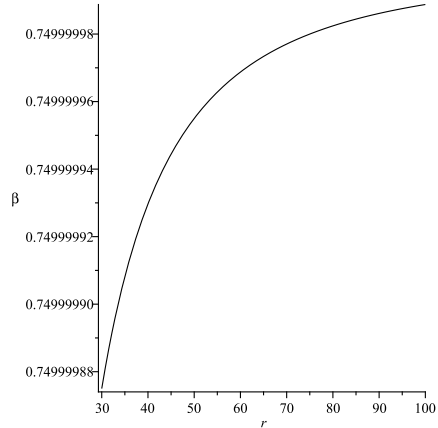}
    \caption{\small Plot for the variation of $\beta = (1 -\frac{p_{t}^2}{p_{r}^2})$ vs $r~(R = 30)$ in galactic halo region.}
    \label{3}
\end{figure}

We note, from the Fig. 3, that the anisotropy in the galactic halo
region gradually decreases with distance. It indicates possible
deviation from spherical symmetry. Dubinsky and Carlberg
\cite{Dubinsky1991} first observed that the halos of elliptic
galaxies are rounder than the shape predicted by cold dark matter
simulation. Recently Kazantzidis et al. \cite{Kazantzidis2004}
reported that cosmological simulations with cooling of baryons in
the galactic halo region are rounder than the shape obtained in
adiabatic simulation. They pointed out that condensation of dark
matter in the galactic halo region due to cooling may result in a
more spherical distribution of dark matter. In the present model
we have taken the galactic halo region dominated by dark matter
which is consistent with the observational evidences. Hence the
predictions of our model is consistent with the cold dark matter
simulations as also mentioned in the Introductory part of the
present paper
\cite{Navaro1996,Navaro1997,Fukushinge1997,Fukushinge2001,Jing2000,Jing2002,Lux2012}.

Let us now find out an expression for the electric field. This can
be given by
\begin{equation}
F =  \frac{\omega}{ R^2 (1+\omega)}\left(3 +
\frac{r^2}{R^2}\right)\left(1 + \frac{r^2}{R^2}\right)^{-2} +
\frac{1}{1 +
 \omega}\left[ \left(1 + \frac{r^2}{R^2}\right)^{-1}\left(\frac{l}{r^2} + \frac{1}{r^2}\right) -
 \frac{1}{r^2}\right].
\end{equation}

\begin{figure}
    \centering
        \includegraphics[scale=.3]{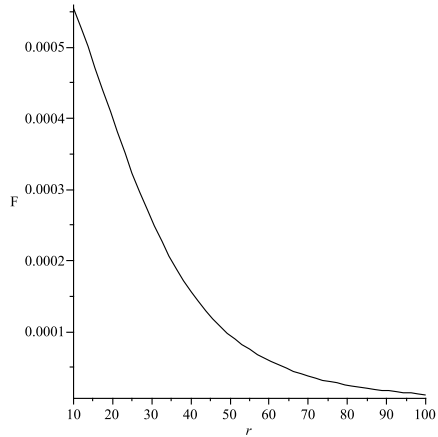}
    \caption{\small Plot for the variation of $F$ vs $r$ in kpc $(R = 40)$ in galactic halo region. }
    \label{4}
\end{figure}

From the Fig. 5 we observe that the electric field in the galactic
halo region decreases gradually with distance from the galactic
center. The distribution of charge in the galactic halo region can
be noted from Fig. 6. The charge density appears to be large near
the galactic center and it decreases steadily with the distance
(Fig. 7).

\section{Stability of orbits of neutral hydrogen clouds}
Let a test particle with four velocity
$U^{\alpha}=\frac{dx^{\sigma}}{d\tau}$ moving in the region of
spacetime metric given in Eq. (\ref{E:line2}). Assuming
$\theta=\pi/2$, the equation
$g_{\nu\sigma}U^{\nu}U^{\sigma}=-m_{0}^{2}$ yields
\begin{equation}
\left( \frac{dr}{d\tau}\right)^{2}=E^{2}+V(r),
\end{equation}
with
\begin{equation}
V(r)=-\left[E^2 \left( 1-\frac{1}{B_0 r^l ( 1 + \frac{r^2}{R^2}
)}\right) +\frac{\left( 1+\frac{L^{2}}{r^{2}}\right)}{\left( 1 +
\frac{r^2}{R^2} \right)} \right].
 \end{equation}

\begin{figure}
    \centering
        \includegraphics[scale=.3]{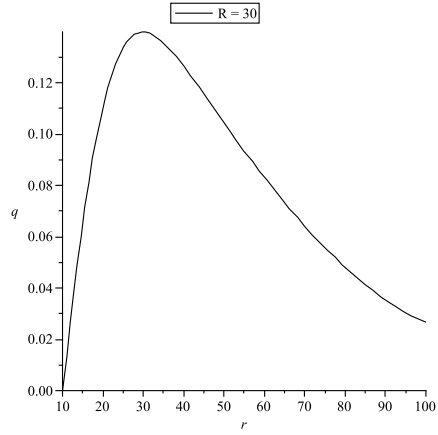}
    \caption{\small Plot for the total charge integrated from $r_{1} = 10$~kpc and $r_{2}$ varied
    from $10$~kpc to $100$~kpc ($R = 30$, $l= 10^{-6}$, $\omega = 1$) in galactic halo region. }
    \label{5}
\end{figure}

\begin{figure}
    \centering
        \includegraphics[scale=.3]{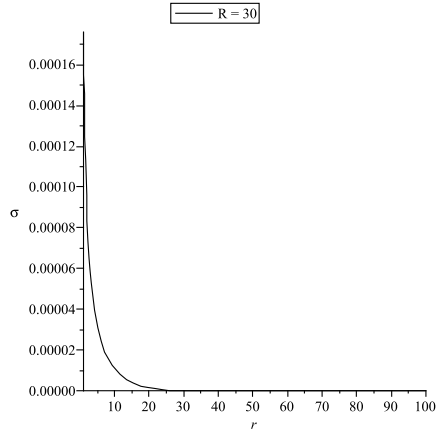}
    \caption{\small Plot for the variation of $\sigma$ vs $r$ in kpc
    ($R = 30$, $l= 10^{-6}$, $\omega = 1$) in galactic halo region. }
    \label{6}
\end{figure}

Here the two conserved quantities, namely relativistic energy
($E$) and angular momentum ($L$) per unit rest mass of the test
particle respectively are
\begin{equation}
\quad E=\frac{U_{0}}{m_{0}}\quad and \quad L=\frac{U_{3}}{m_{0}}.
\end{equation}

If the circular orbits are defined by $r=a$, then
$\frac{da_0}{d\tau}=0$ and, additionally,
$\frac{dV}{dr}\mid_{r=a}=0$. Above two conditions result
\begin{equation}
L=\pm\sqrt{\frac{l}{2-l}}a,
\end{equation}
and by using $L$ in $V(a)=-E^{2}$, we get
\begin{equation}
E=\pm\sqrt{\frac{2B_{0}}{2-l}}a^{l/2}.
\end{equation}
The orbits will be stable if $\frac{d^{2}V}{dr^{2}}\mid_{r=a}<0$
and unstable if $\frac{d^{2}V}{dr^{2}}\mid_{r=a}>0$.

By putting the expressions for $L$ and $E$ in
$\frac{d^{2}V}{dr^{2}}\mid_{r=a}$, finally we get
\begin{eqnarray}
\frac{d^{2}V}{dr^{2}}\mid_{r=a}=- \frac{R^{2}}{B_{0} \left( R^{2}
+ a^{2}\right)^{3} a^{4}} \left( -E^{2}a^{(2 - l)} l^{2} R^{4} -2
E^{2} a^{4-l} l^{2}R^{2} - E^{2} a^{(6-l)} l^{2} \right. \nonumber
\\ - \left.  6E^{2}a^{(4-l)}lR^{2} - 5E^{2}a^{(6-l)}l - E^{2}a^{(2-l)}lR^{4}
6E^{2}a^{(6-l)}+2E^{2}a^{(4-l)}R^{2} \right. \nonumber
\\ + \left. 6L^{2}B_{0}R^{4} +
18L^{2}B_{0}R^{2}a^{2} + 20L^{2}B_{0}a^{4} + 6a^{6}B_{0} -
2B_{0}a^{4}R^{2} \right).
\end{eqnarray}

The second derivative of the gravitational potential function is
plotted for different values of the parameters $E$, $l$, $R$,
$B_{0}$ and $L$. The plots are given in Figs.(8), (9) and (10).
The plots show that the second derivative is negative for
arbitrary ranges of $r$. So the orbits of neutral hydrogen cloud
must be stable in galactic halo region. The plots reveal an
important relationship between angular momentum per unit mass of
the test particles and the stability of the orbits. For $L = 1$
(Fig. 9) the orbits are unstable up to the distance of nearly $10$
kpc. However, for comparatively large values of $L$, the orbits
are stable in the range from $10$~kpc to $300$~kpc.

\begin{figure}
    \centering
        \includegraphics[scale=.3]{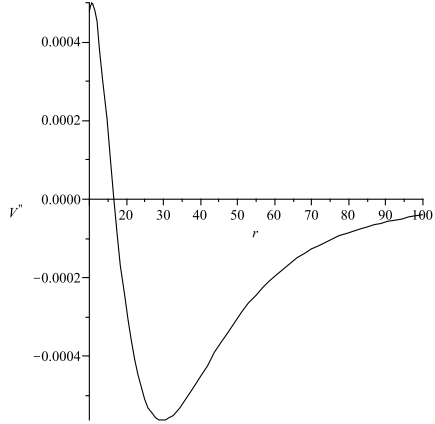}
    \caption{\small Plot for the variation of $V^{\prime \prime}$
    vs $r$ [$ L = 1$, $R = 30$, $l = 10^{-6}$, $E = 10^{-4}$, $B_{0}=100$]. }
    \label{6}
\end{figure}

\begin{figure}
    \centering
        \includegraphics[scale=.3]{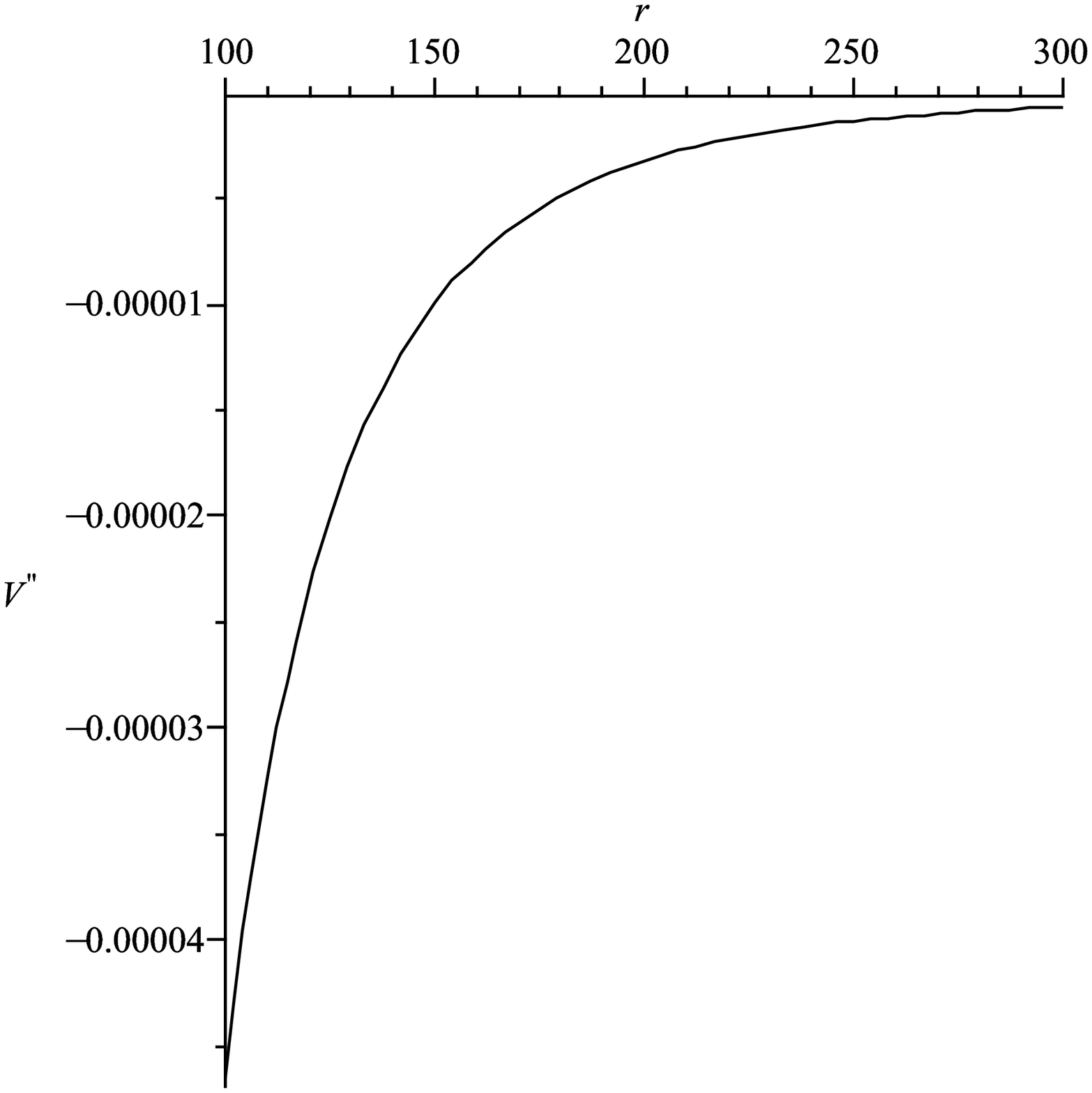}
    \caption{\small Plot for the variation of $V^{\prime \prime}$
    vs $r$ [$ L = 20$, $R = 30$, $l = 10^{-6}$, $E = 10^{-4}$, $B_{0}=100$]. }
    \label{7}
\end{figure}

\begin{figure}
    \centering
        \includegraphics[scale=.3]{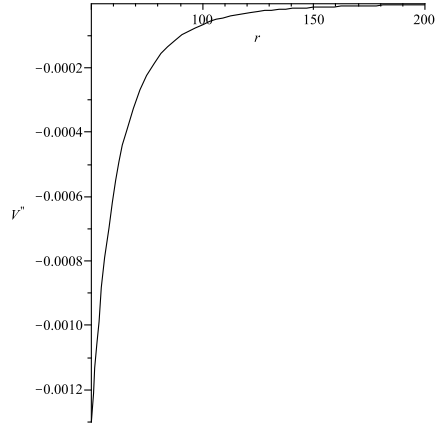}
    \caption{\small Plot for the variation of $V^{\prime \prime}$
    vs $r$ [$ L = 40$, $R = 30$, $l = 10^{-6}$, $E = 10^{-4}$, $B_{0}=100$]. }
    \label{8}
\end{figure}

\section{Motion of test particles in galactic halo region}
Since the observations indicate that gravity in the galactic scale
would be attractive, therefore, it is interesting to check whether
the particles are being accelerated towards the galactic center.
Geodesic study will help to see this important characteristic.

Now we study the geodesic equation for a test particle that has
been ``placed'' at some radius $r_{0}$ and is given by
\begin{equation}
\frac{d^{2}x^{\alpha}}{d\tau^{2}}+\Gamma_{\alpha}^{\mu\gamma}\frac{dx^{\mu}%
}{d\tau}\frac{dx^{\gamma}}{d\tau}=0.
\end{equation}

This yields the radial equation
\begin{equation}
\frac{d^2 r}{d\tau^2}= -\frac{1}{2} \left( 1 + \frac{r^2}{R^2}
\right)^{-1} \left[\frac{d}{dr}\left( 1 + \frac{r^2}{R^2} \right)
 \left( \frac{dr}{d\tau}\right)^2 + \frac{d}{dr}B_0r^l \left( \frac{dt}{d\tau}\right)^2 \right]. \label{atrract}
\end{equation}

Since $\left( \frac{dr}{d\tau}\right)_{ r= r_0} = 0$ for a stable
orbit, we find from equation (\ref{atrract})
\begin{equation}
\frac{d^{2}r}{d\tau^{2}}= - \frac{1}{2} \frac{1}{\left(1 +
\frac{r^2}{R^2}\right)} B_{0} l r^{l-1}
\left(\frac{dt}{d\tau}\right)^2 < 0.
\end{equation}

Thus particles are attracted towards the center. Interestingly,
since we are considering the rotation of neutral hydrogen clouds,
the electric field will have no direct effect on the motion.
However, the electric field contributes to the energy-momentum
tensor of the system which in turn influences the geometry of the
space-time. Thus, the electric field indirectly affects the motion
of the neutral hydrogen clouds.

\section{The total gravitational energy}
We try to find out the total gravitational energy $E_{G}$ between
two fixed radii, say, $r_{1}$ and $r_{2}$ which is given by
\[E_G = M - E_M = 4\pi \int_{r_1}^{r_2}\left[1-\sqrt{\left(1 -
\frac{r^2}{R^2}\right)}\right]\rho r^2 dr.\]

This study will be interesting as positive energy density does not
always lead to attractive gravity \cite{Nandi2009}. According to
Misner, Thorne and Wheeler \cite{Misner1973} and Nandi et al.
\cite{Nandi2009} the total gravitational energy in the halo region
should be negative.

Now, plugging $\rho$ from Eq. (\ref{E:rho}), we get for the total
gravitational energy $E_{G}$ as follows:
\begin{eqnarray}
E_G=4\pi\int_{r_{1}}^{r_{2}}\left[1 - \sqrt{\left(1 -
\frac{r^2}{R^2}\right)}\right] \times \nonumber \\
\left[\frac{1}{8\pi(1+\omega)}\left\{\frac{1}{R^2}\left(3 +
\frac{r^2}{R^2}\right) \left(1 + \frac{r^2}{R^2}\right)^{-2} +
\left(1 + \frac{r^2}{R^2}\right)^{-1}\left(\frac{l+1}{r^2}\right)
- \frac{1}{r^2}\right\}\right]r^2dr,
\end{eqnarray}
where

\begin{eqnarray}
M=\left(\frac{1}{2(1+\omega)}\right)\int_{r_1}^{r_2}\left[\frac{1}{R^2}\left(3
+ \frac{r^2}{R^2}\right) \left(1 + \frac{r^2}{R^2}\right)^{-2}
\right. \nonumber
\\ + \left.
\left(1 + \frac{r^2}{R^2}\right)^{-1}\left(\frac{l+1}{r^2}\right)
- \frac{1}{r^2}\right]r^2dr,
\end{eqnarray}
i.e. the Newtonian mass can be expressed as
\begin{equation}
 M =\frac{1}{8(1+ \omega)} \left[ R(l+1)\tan^{-1}\left(\frac{r}{R}\right) - R^{2}\left( \frac{r}{R^2 +
r^2}\right)\right]_{r{1}}^{r{2}}.
\end{equation}

Thus we get the total gravitational energy as follows:
\begin{eqnarray}
 E_{G}   = \frac{1}{2 ( 1 + \omega )}\left[R(l + 1)\tan^{-1}\left(\frac{r}{R}\right)
 - \frac{R^2r}{r^2 + R^2} - R(l + 2)\log \vert r + \sqrt{R^2 + r^2}\vert \right.
\nonumber
\\ + \left.
  \frac{2Rr}{ \sqrt{R^2 + r^2}}\right]_{r_{1}}^{r_{2}}.
\end{eqnarray}

Figs. (11) and (12) show that the total gravitational energy in
the galactic halo region is negative for arbitrary choice of
positive values of $r_{1}$ and $r_{2}$. It may be noted this is
consistent with the present day observations.

\begin{figure}
    \centering
        \includegraphics[scale=.3]{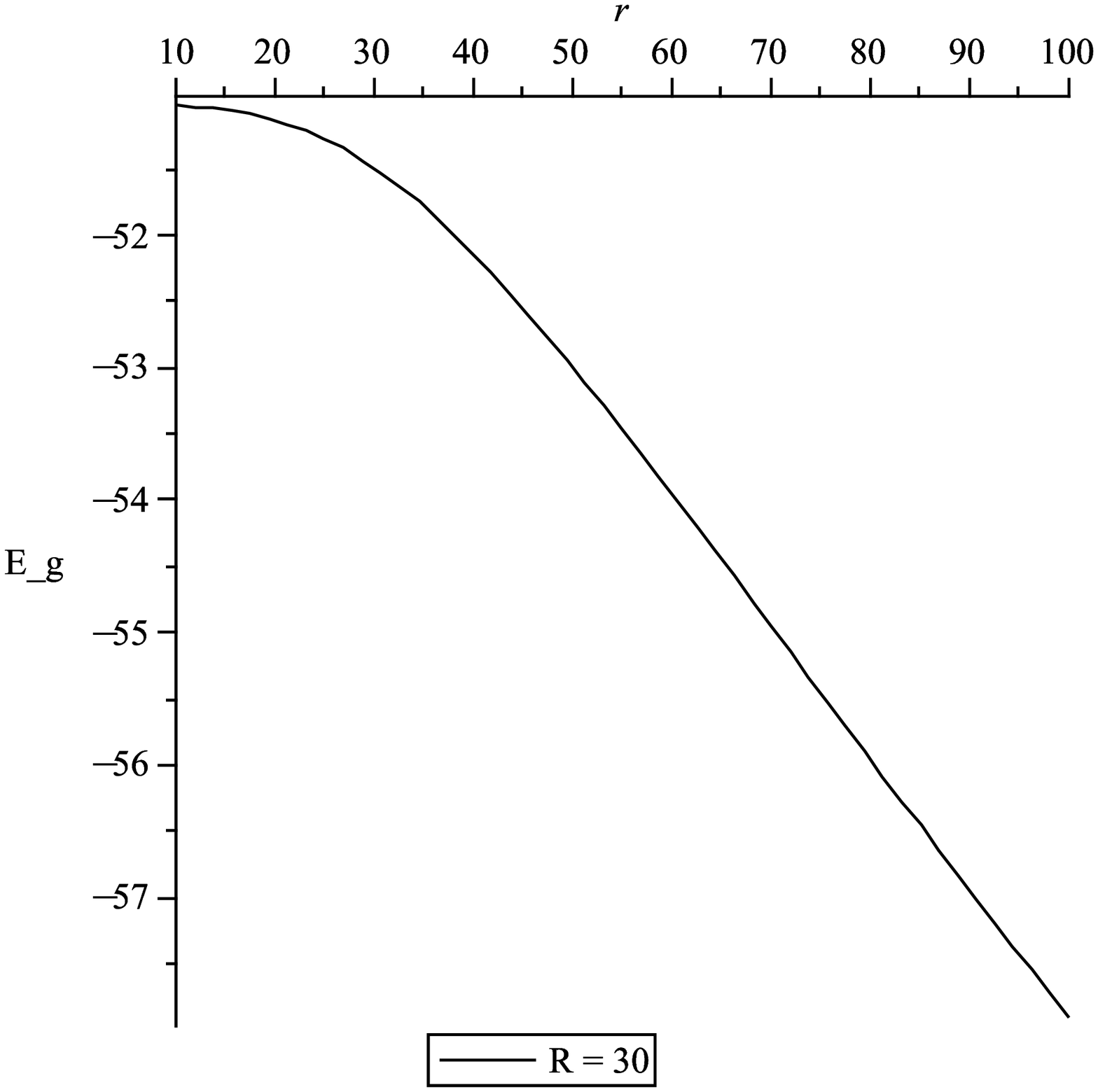}
    \caption{\small Plot for the total gravitational energy $E_{g}$ vs $r$ in Kpc. }
    \label{9}
\end{figure}

\begin{figure}
    \centering
        \includegraphics[scale=.3]{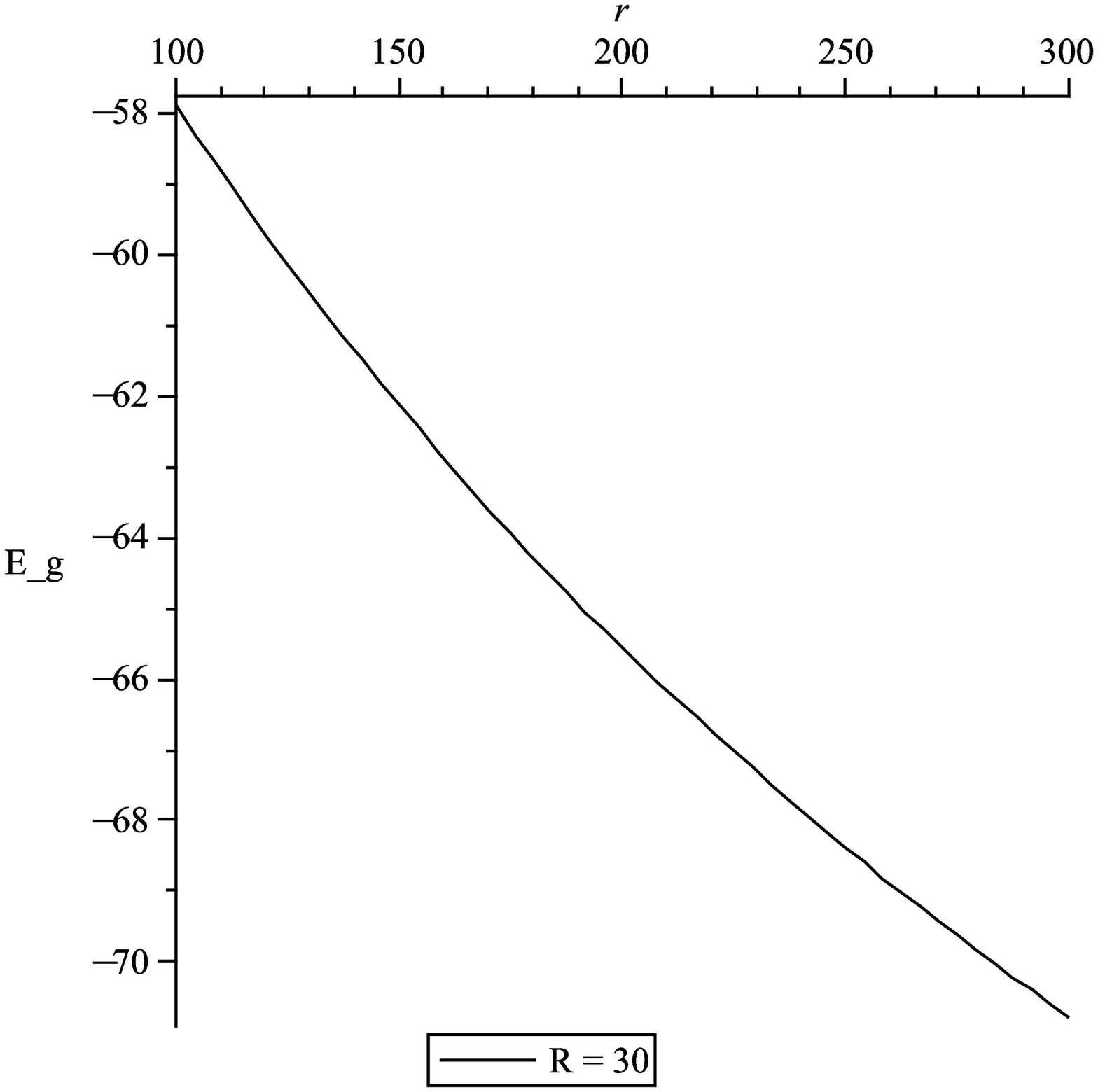}
    \caption{\small Plot for the total gravitational energy $E_{g}$ vs $r$ in Kpc. }
    \label{9}
\end{figure}

\section{The observed equation of state}
From the gravitational lensing measurements and the gravitational
potential and total mass-energy inferred from the rotational
curves we can approximately obtain a dimensionless parameter which
gives the rough measure of the equation of state of the fluid in
the galactic halo region \cite{Faber2006}. But first of all let us
rewrite the metric of static, spherically symmetric spacetime as
given in Eq.~(\ref{line1}), in the following form:
\begin{equation}
ds^2 = - e^{2\phi(r)} dt^2
                   + \frac{1}{1 - 2 m(r)/r}dr^2+r^2 d\Omega_2^2,  \label{line3}
\end{equation}
where $m(r)$ and $\phi(r)$ are two functions which determine the
metric uniquely. Comparing the above metric (\ref{line3}) with the
metric (\ref{line1}), we get
\begin{equation}
\phi(r) = \frac{1}{2} \left[ \ln B_{0} + l \ln
r\right],
\end{equation}
and
\begin{equation}
m(r) = \frac{1}{2} \left[ \frac{r^{3}}{R^{2} + r^{2}}\right].
\end{equation}

From the Einstein's field equations for the metric (\ref{line3}),
$m(r)$ can be interpreted as the total mass-energy within a
spherical distribution of radius $r$ and $\phi(r)$ is the
gravitational potential. From the Fig. 13 we can note that
function $m(r)$ increases with $r$. The plot resembles the plot of
cumulative mass versus radius in the paper of van Albada et al.
\cite{Albada1985}.

\begin{figure}
    \centering
        \includegraphics[scale=.3]{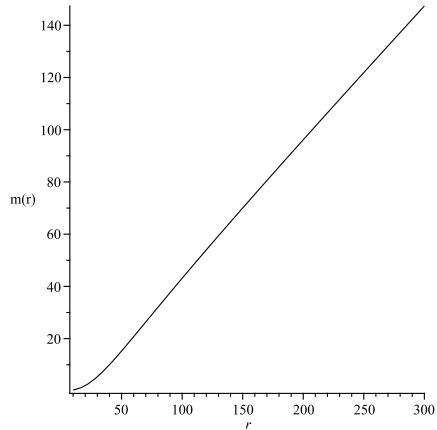}
    \caption{\small Plot for the variation of $m(r)$ vs $r$ in Kpc.}
    \label{10}
\end{figure}

Now, let us define $\Phi(r)_{RC}$ as the gravitational potential
obtained from the measurements of rotational curves. Since the
test particles moving in the galactic halo region are governed by
the general relativistic potential $\Phi(r)$, therefore one can
conclude $\Phi(r) = \Phi_{RC}(r)$. Similarly, $m_{RC}(r)$ is
defined as the mass obtained from the rotational curve
measurements. As given by Faber and Visser \cite{Faber2006}, the
values of these functions are:
\begin{equation}
\phi_{RC} = \phi(r)= \frac{l}{2r},
\end{equation}
and
\begin{equation}
m_{RC} = r^2\Phi^{\prime}(r)=\frac{1}{2}lr.
\end{equation}

The primes denote the derivatives with respect to $r$. The
quantity $m_{RC}$ represents the mass estimate from the rotational
curves. It is interesting to note from the aforementioned equation
that $m_{RC} \propto r$ (Fig. 14). This is in conformity with the
observational evidences.

\begin{figure}
    \centering
        \includegraphics[scale=.3]{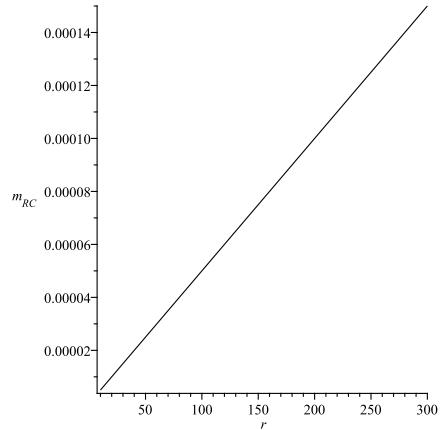}
    \caption{\small Plot for the variation of $m_{RC}$ vs $r$ in Kpc.}
    \label{11}
\end{figure}

As discussed in Nandi et al. \cite{Nandi2009}, the functions are
determined indirectly from certain lensing measurements defined by
\begin{equation}
\Phi_{lens}=\frac{\Phi(r)}{2}+\frac{1}{2}\int\frac{m(r)}{r^{2}}dr
= \frac{\ln B_{0}}{4} + \frac{l \ln r}{4} + \frac{\ln (r^2 +
R^2)}{8},
\end{equation}
and
\begin{equation}
m_{lens}=\frac{1}{2}r^{2}\Phi^{\prime}(r)+\frac{1}{2}m(r) =
\frac{lr}{4} + \frac{r^{3}}{4\left( R^{2} + r^{2} \right)}.
\end{equation}

Of particular interest to us is the dimensionless quantity
\begin{equation}
\omega(r)=
\frac{2}{3}\frac{m_{RC}^{\prime}-m_{lens}^{\prime}}{2m_{lens}^{\prime}-m_{RC}^{\prime}},
\end{equation}
due to Faber and Visser \cite{Faber2006}. In the above expression
the subscript $RC$ refers to the rotation curve. Thus, the
resulting expression is
\begin{equation}
\omega(r)=
\frac{2}{3}\frac{m_{RC}^{\prime}-m_{lens}^{\prime}}{2m_{lens}^{\prime}-m_{RC}^{\prime}}
= \frac{l\left( R^{2} + r^{2}\right)^{2}}{3\left( 3r^{2}R^{2} +
r^{4}\right)}.
\end{equation}

\begin{figure}
    \centering
        \includegraphics[scale=.3]{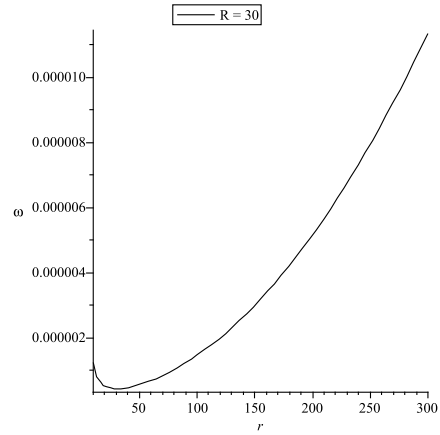}
    \caption{\small Variation of observed equation of state parameter $\omega(r)$ with $r$ in kpc. $R = 30$}
    \label{12}
\end{figure}

\begin{figure}
    \centering
        \includegraphics[scale=.3]{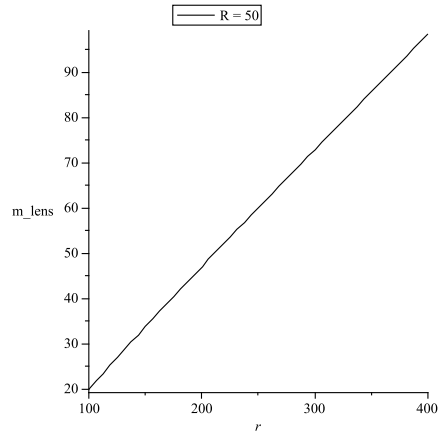}
    \caption{\small Variation of $m_{lens}$ with $r$ in kpc. $R = 50$, $B_{0}= 100$ }
    \label{13}
\end{figure}

As we can note from the Fig. 15 that the observed equation of
state parameter in the galactic halo region is always positive.
This indicates that the matter content in the galactic halo region
must be normal matter. This is in confirmation with the earlier
works of Rahaman et al. \cite{Rahaman2010}.

On the other hand, small variation of observed equation of state
parameter with distance is shown in the plot (Fig. 16). The
parameter $m_{lens}$ represents pseudo mass in the galactic halo
region contributing to lensing potential, $\Phi_{lens}$
\cite{Faber2006}. The Fig. 16 shows steady increase in $m_{lens}$
in the galactic halo region $\left(m_{lens} \propto r \right)$. We
should remember that $m_{lens}$ is the mass to be observed in the
halo region. The observations also predict linear increase in mass
in the galactic halo $\left( M(r) \propto r \right)$. Thus the
predictions of the model is supported by observational evidences.

\section{Post Newtonian mass in galactic halo region}
In Newtonian regime the pressures of galactic fluids are to be
negligible \cite{Faber2006}. We have calculated the Newtonian mass
of the galactic fluid in Eq. (30). The post Newtonian mass of the
galactic fluid is given by the following equation
\[M_{pN}=4\pi\int_{r_{1}}^{r_{2}}(\rho+p_r+2p_t) r^{2}dr. \]

Thus, after plugging the parameters $\rho$, $p_r$, $p_t$ and
performing integration, we get the post Newtonian mass of the
galactic fluid in the following form:
\begin{eqnarray}
 M_{pN}= \left[\frac{1}{2} \left\{ R (l +1) \tan^{-1}\left(\frac{r}{R}\right) -
\frac{rR^{2}}{r^2+R^2}\right\} + \left( \frac{l^2-l-2}{4} -
\frac{l+1}{\omega + 1}   \right)
 R\tan^{-1}\left(\frac{r}{R}\right
 ) \right. \nonumber
\\ + \left. \left(   \frac{l+2}{4} +
\frac{\omega}{1 + \omega} \right) \frac{R^{2}r^{2}}{R^2 + r^2}-
 \frac{\omega r}{1+ \omega} \right]_{r_{1}}^{r_{2}}.
\end{eqnarray}

As can be noted from the plots of Newtonian and post Newtonian
mass (Figs. 16-19), both increases linearly with distance from
$100$ kpc to $300$ kpc. The Figs. 17 and 18 show that the
Newtonian mass increases nonlinearly up to $60$ kpc from the
galactic center. Beyond that the increase becomes linear with
distance.

\begin{figure}
    \centering
        \includegraphics[scale=.3]{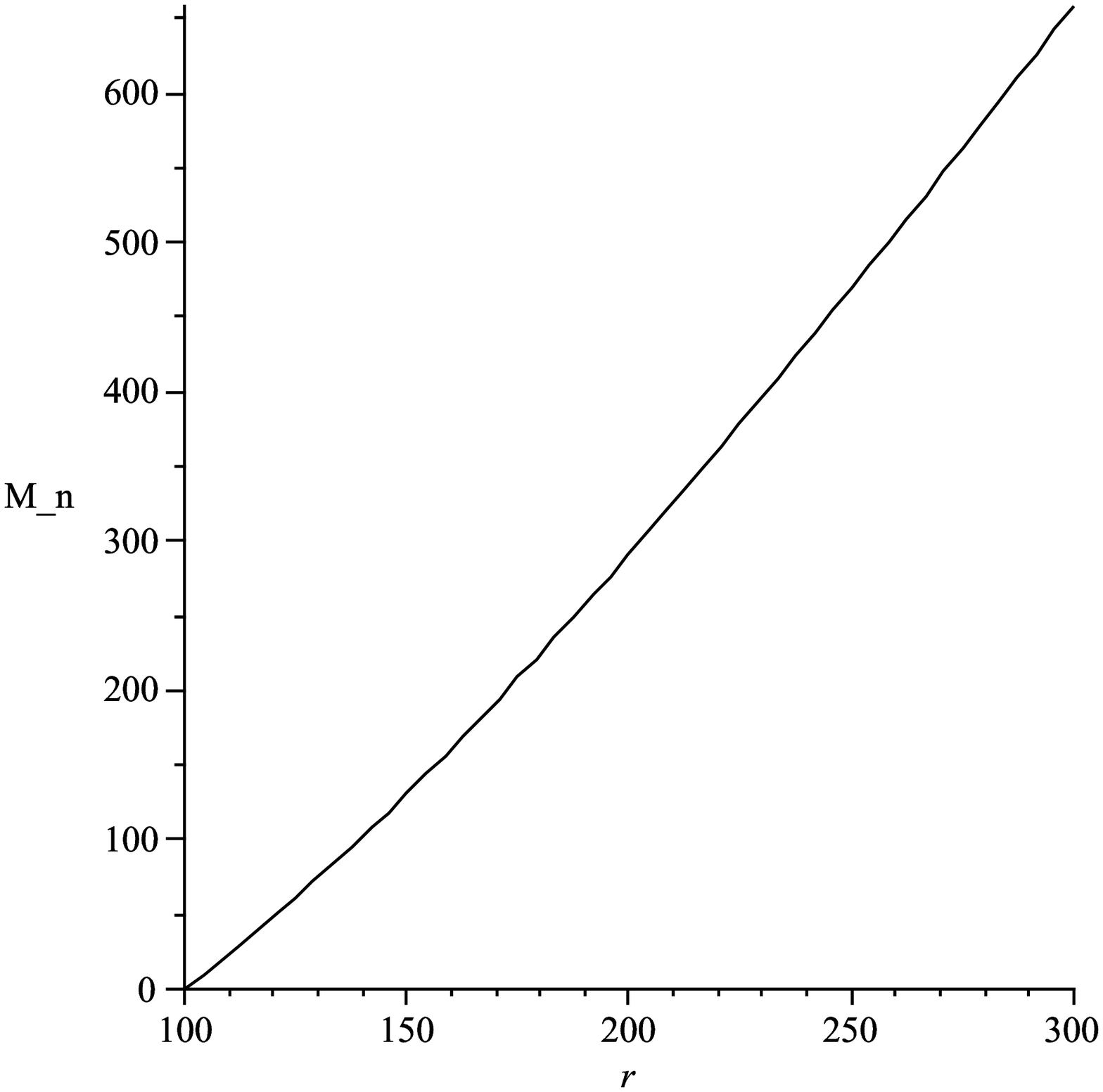}
    \caption{\small Plot for the Newtonian mass $M_{N}$ vs $r$ (in kpc)
    in galactic halo region with different range. }
    \label{16}
\end{figure}

\begin{figure}
    \centering
        \includegraphics[scale=.3]{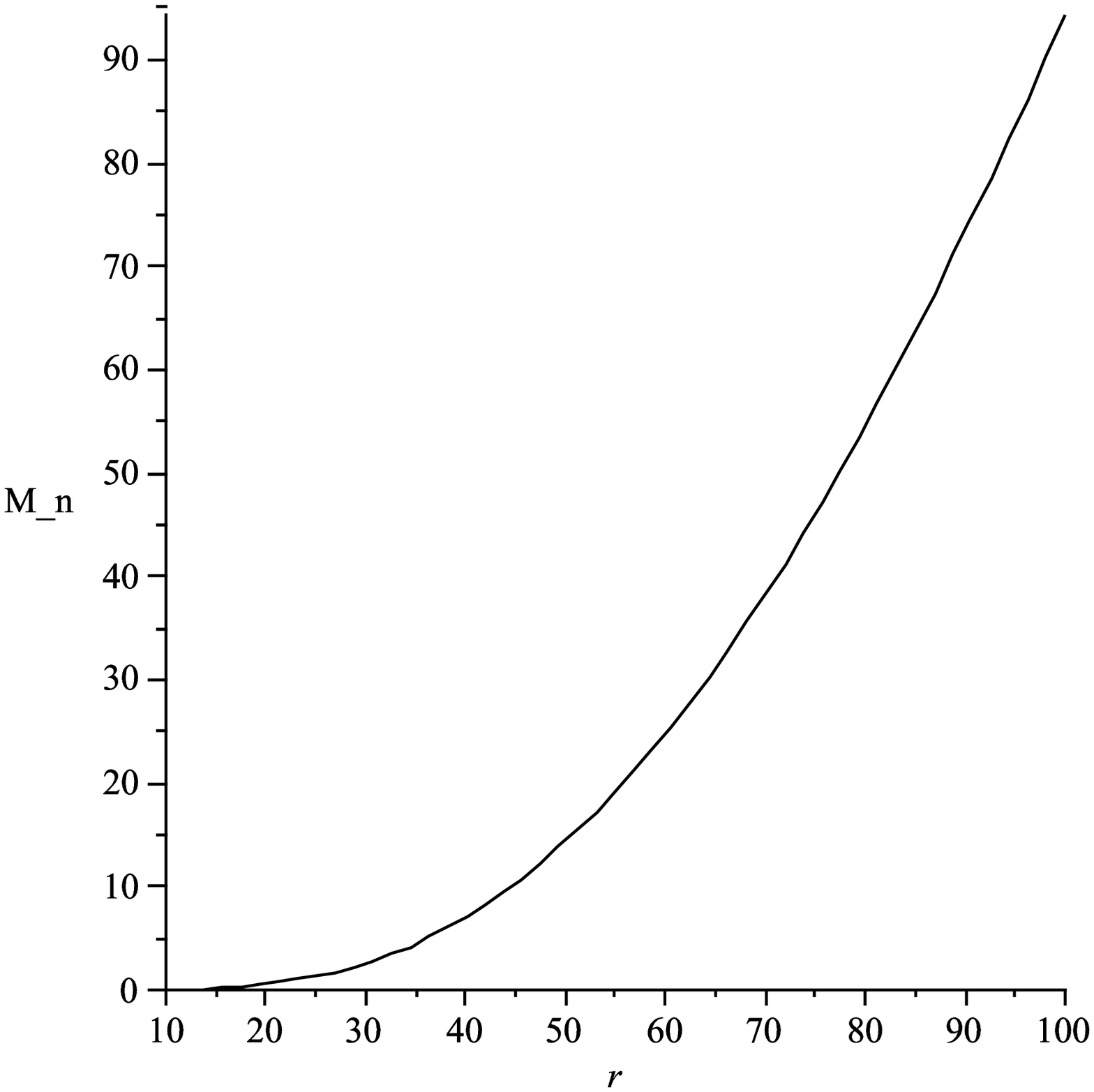}
    \caption{\small Plot for the Newtonian mass $M_{N}$ vs $r$ (in kpc)
    in galactic halo region with different range. }
    \label{17}
\end{figure}

On the other hand, the increase of post Newtonian mass becomes
linear from $30$ kpc from galactic center (Figs. 19 and 20). This
implies that in the halo region Newtonian as well as post
Newtonian mass is proportional to $r$. The observational
predictions from the rotational curves in the galactic halo region
are also in accordance with this result. From numerical simulation
we found that the ratio of post Newtonian to Newtonian mass of
galactic fluid in the range $100$~kpc to $300$~kpc is nearly
$659.7984402$. In the range $10$~kpc to $100$~kpc the post
Newtonian mass is nearly $1125.448032$ times larger than Newtonian
mass. Since the post Newtonian mass indicates the relativistic
effect, one can note from the figures that relativistic effects
are considerable in the galactic halo region.

\begin{figure}
    \centering
        \includegraphics[scale=.3]{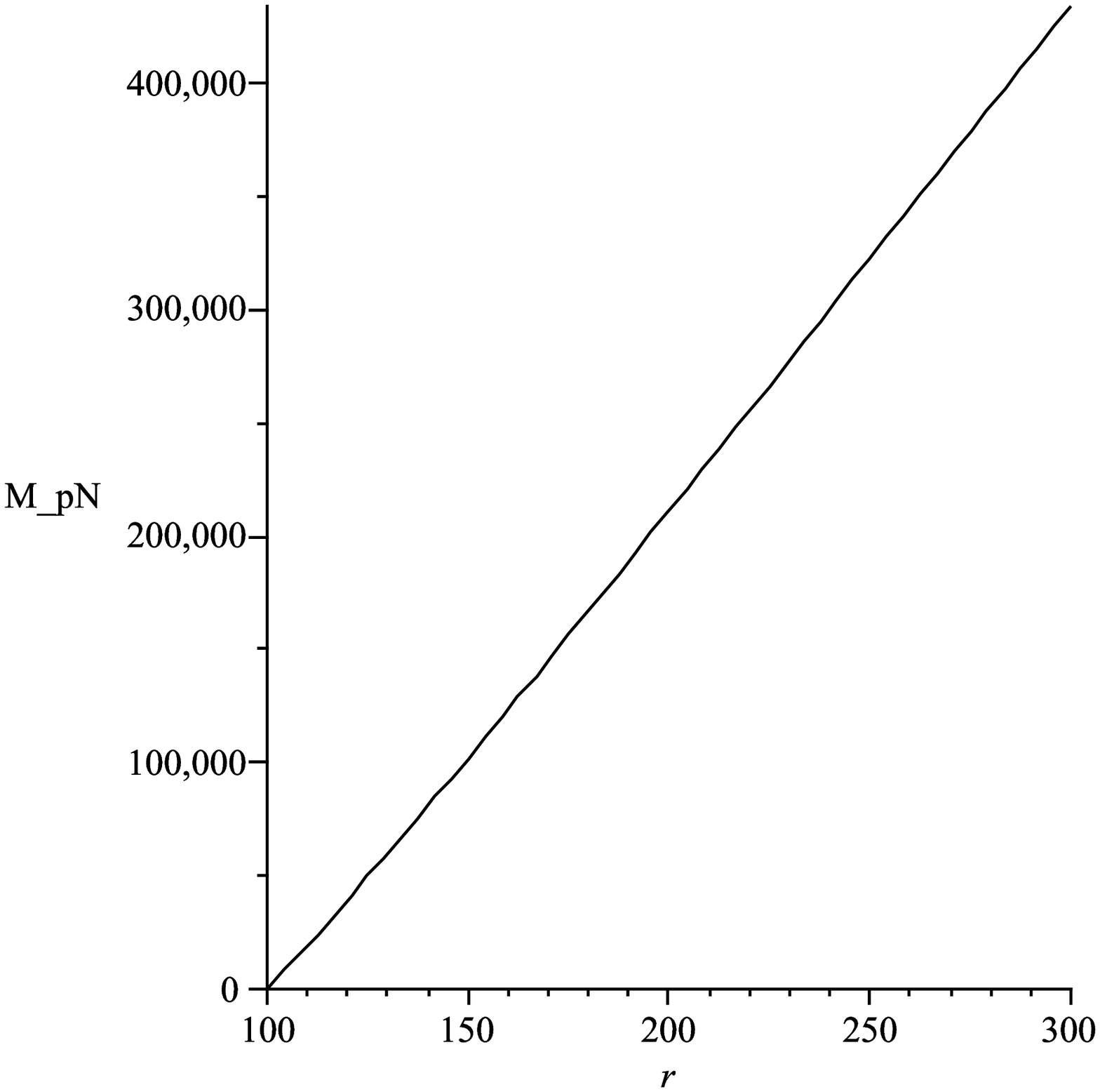}
    \caption{\small Plot for the post Newtonian mass $M_{pN}$ vs $r$ (in kpc)
    in galactic halo region with different range. }
    \label{18}
\end{figure}
\begin{figure}
    \centering
        \includegraphics[scale=.3]{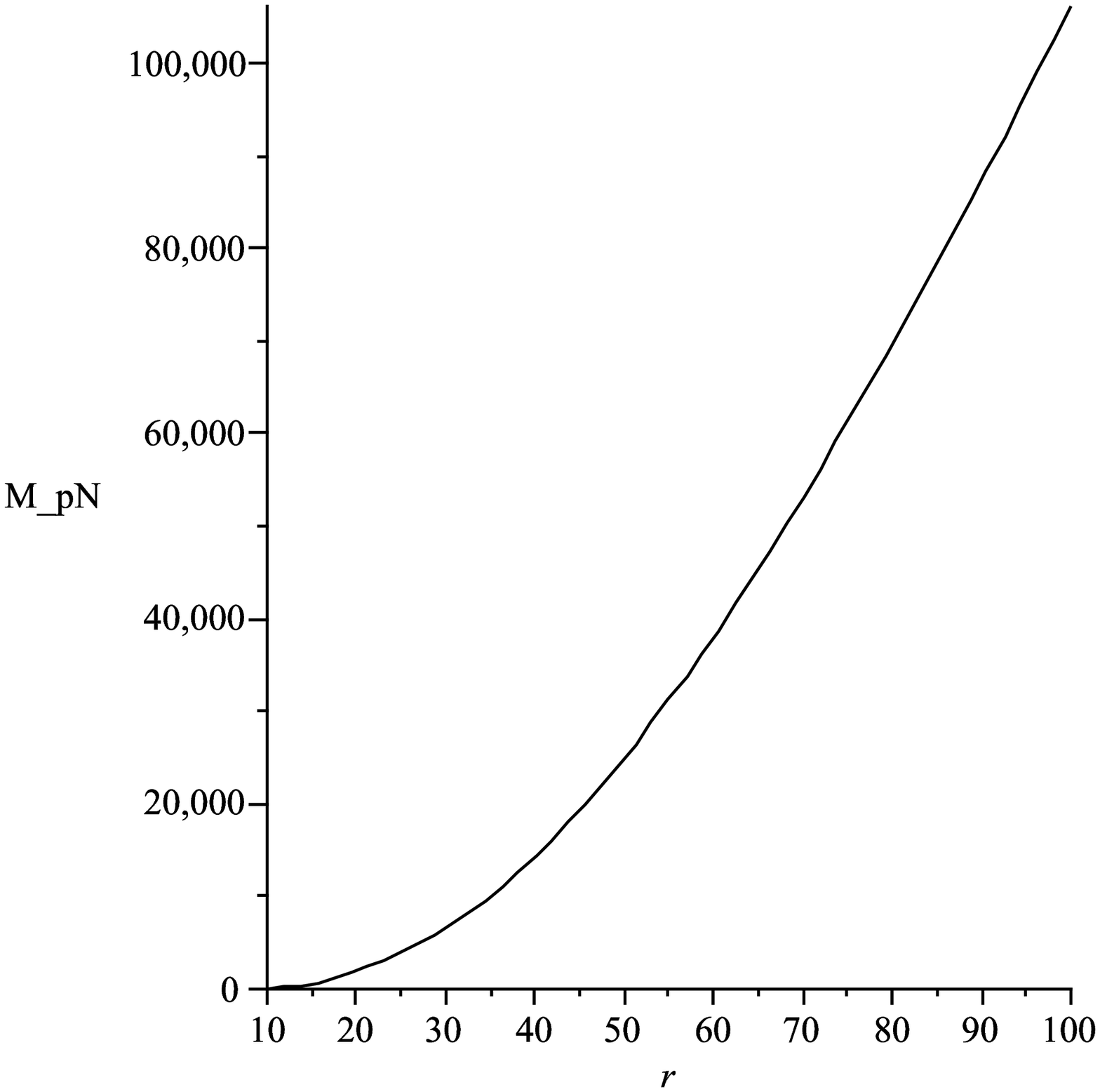}
    \caption{\small Plot for the post Newtonian mass $M_{pN}$ vs $r$ (in kpc)
    in galactic halo region with different range. }
    \label{19}
\end{figure}

\section{Concluding Remarks}
In the foregoing, we have proposed a model of galactic halo with
electric field. The motivation behind such proposal is that only
Einstein's General Theory of Relativity fail to explain the
observed phenomena of the rotation curves of spiral galaxies.
Researchers need to hypothesize the existence of some non luminous
dark matter of unknown properties to be responsible for this
peculiar phenomena. The assumption that `mass increases with
radial distance' may support the observed result, however,
luminous mass distribution does not follow this behavior. We  have
proposed the model of galactic halo with electric field rather
considering the unknown nature of the dark matter.

Different cosmological models (e.g. SCDM, $\Lambda$-CDM etc.)
predict attractive gravity in galactic halo region and beyond. We
have confirmed the existence of attractive gravity in the halo
region (Sec. 5). The orbits of revolving particles are found to be
stable in the halo. However, in the region within $10$~kpc from
galactic center, the orbits are unstable for particles of
comparatively small angular momentum per unit mass. For Milky way
galaxy this region belongs to the stellar halo region.

The geometry of the halo region is calculated from the
observational constraint. The results of the present paper show
that being considered as a fluid, possibly, dark matter is not
exotic in nature. However, from the point of view of particle
physics what dark matter is constituted of is yet to be resolved.
Our results has nothing to remark on the particle aspect of the
dark matter. The pressure of dark matter as predicted in SCDM
paradigm is negligible. We have confirmed this theoretical
prediction in our results. The observational features predicted
here (Sec. 6)  are also in agreement with the present
observational evidences regarding galactic halo.

The present model indicates that in the halo region Newtonian as
well as post Newtonian mass is proportional to $r$. The
observational predictions from the rotational curves in the
galactic halo region are also in accordance with this result. From
numerical simulation we found that the ratio of post Newtonian to
Newtonian mass of galactic fluid in the range $100$~kpc to
$300$~kpc is nearly $659$. Though the post Newtonian mass
indicates the relativistic effect, taking into consideration the
large distance scale in the galactic halo, one can note that
relativistic effects will not be appreciably high in the galactic
halo region. Thus interestingly, we find in our model that $M_{pN}
\approx 659 M(r)$ in contrast to the well known scalar field model
$M_{pN} = 10^{6} M(r)$ \cite{Matos2000}. As per observational
evidences, the interaction cross-section of dark matter with
normal baryonic matter must be very small. Commonly referred dark
matter particle candidates like WIMP, axions etc. also supports
the prediction \cite{Overduin2004}. Since we have taken the
baryonic matter content in the galactic halo region negligible
compared to the dark matter, the present model predicts that the
role of electric field in anisotropy of galactic halo must be very
weak.

A proper relativistic model of galactic halo with electric field
must be based on electrohydrodynamics. However, ours is a simpler
model in which we have assumed the motion of charged particles
within the galactic halo insignificant. Also, the magnetic field
is assumed much less than the electric field. Commonly, magnetic
field is regarded as significant in halos. However, with the
assumptions aforementioned, it is shown that the model predicts
successfully some of the observational features regarding galactic
halo. This model can be extended farther with the help of
electro-hydrodynamics and electromagneto-hydrodynamics.

The present work may be extended by considering magnetic field
along with the electric field. A more realistic model of galactic
halo may be thought of with the superposition of halo with
galactic disc \cite{vogt1,lora,gutierrez1} or central black hole
\cite{semerak,vogt2,gutierrez2}.

\section*{Acknowledgments}
KC, FR and SR are thankful to the authority of Inter-University
Centre for Astronomy and Astrophysics, Pune, India for providing
them Visiting Associateship under which a part of this work was
carried out. FR is also thankful to UGC, for providing financial
support under research award scheme. KC is thankful to UGC for
providing financial support in MRP under which this research work
was carried out. We all are grateful to the anonymous referees for
their constructive suggestions which have enabled us to upgrade
the manuscript substantially.

\end{document}